\documentclass[%
reprint,
amsmath,amssymb,
aps,
prb,
]{revtex4-1}

\usepackage[utf8]{inputenc}
\usepackage{xcolor}
\usepackage{graphicx}
\usepackage{dcolumn}
\usepackage{bm}
\usepackage{mathrsfs}
\usepackage{hyperref}
\usepackage[normalem]{ulem}
\usepackage{lipsum}
\usepackage{IEEEtrantools}
\usepackage[version=3]{mhchem}
\usepackage{verbatim}
\usepackage{siunitx}
\usepackage{enumitem}
\usepackage{textcomp}
\usepackage{lineno}

\definecolor{turq}{rgb}{0.0, 0.72, 0.92}

\setlist[itemize]{noitemsep,topsep=-0.2cm}

\begin{document}
\title{Shared Metadata for Data-Centric Materials Science}

\author{Luca M. Ghiringhelli$^{1,2}$,
Carsten Baldauf$^3$,
Tristan Bereau$^4$, 
Sandor Brockhauser$^1$, 
Christian Carbogno$^2$, 
Javad Chamanara$^{5}$, 
Stefano Cozzini$^6$, 
Stefano Curtarolo$^7$, 
Claudia Draxl$^{1,2}$, 
Shyam Dwaraknath$^{8}$, 
Ádám Fekete$^1$, 
James Kermode$^{9}$, 
Christoph T. Koch$^1$, 
Markus Kühbach$^1$, 
Alvin Noe Ladines$^1$,
Patrick Lambrix$^{10}$, 
Maja-Olivia Lenz-Himmer$^2$,
Sergey Levchenko$^2$, 
Micael Oliveira$^{11}$, 
Adam Michalchuk$^{12}$, 
Ron Miller$^{13}$, 
Berk Onat$^{9}$,
Pasquale Pavone$^1$, 
Giovanni Pizzi$^{14}$, 
Benjamin Regler$^2$, 
Gian-Marco Rignanese$^{15}$, 
Jörg Schaarschmidt $^{16}$,
Markus Scheidgen$^1$,
Astrid Schneidewind$^{17}$,
Tatyana Sheveleva$^{5}$, 
Chuanxun Su$^{18}$, 
Denis Usvyat$^{19}$,
Omar Valsson$^{20}$, 
Christof W\"{o}ll$^{21}$,
and Matthias Scheffler$^{1,2}$\\ 
\vspace{12pt}
\textit{$^1$Physics Department and IRIS Adlershof, Humboldt-Universität zu Berlin, Germany; $^2$The \mbox{NOMAD} Lab at the Fritz Haber Institute of the Max Planck Society and Humboldt University, Berlin, Germany; $^3$Fritz Haber Institute of the Max Planck Society, Berlin, Germany; $^4$Van 't Hoff Institute for Molecular Sciences and Informatics Institute, University of Amsterdam, Amsterdam 1098 XH, The Netherlands; $^{5}$TIB – Leibniz Information Centre for Science and Technology and University Library, Welfengarten 1B, 30167 Hanover, Germany; $^6$AREA Science Park, località Padriciano, 34149 Trieste, Italy; $^7$Center for Autonomous Materials Design and Department of Mechanical Engineering and Materials Science, Duke University, Durham NC 27708, USA; $^{8}$Lawrence Berkeley National Laboratory, Berkeley, CA, USA; $^{9}$Warwick Centre for Predictive Modelling, School of Engineering, University of Warwick, Coventry, CV4 7AL, United Kingdom; $^{10}$Department of Computer and Information Science, Linköping University, Sweden and The Swedish e-Science Research Centre, Linköping University,Sweden; $^{11}$Max Planck Institute for the Structure and Dynamics of Matter, Hamburg, Germany; $^{12}$Federal Institute for Materials Research and Testing (BAM), 12489 Berlin, Germany; $^{13}$Department of Mechanical and Aerospace Engineering, Carleton University, Ottawa, ON K1S 5B6, Canada; ${14}$Theory and Simulation of Materials (THEOS) and National Centre for Computational Design and Discovery of Novel Materials (MARVEL), École Polytechnique Fédérale de Lausanne, CH-1015 Lausanne, Switzerland; $^{15}$Institute of Condensed Matter and Nanosciences (IMCN), UCLouvain, Chemin des Étoiles 8, B-1348 Louvain-la-Neuve, Belgium; $^{16}$Institute of Nanotechnology, Karlsruhe Institute of Technology (KIT), 76344 Eggenstein-Leopoldshafen, Germany; $^{17}$ Jülich Center for Neutron Science at MLZ, Forschungszentrum Jülich GmbH, Lichtenbergstraße 1, 85748 Garching, Germany; $^{18}$CAS Key Laboratory of Quantum Information, University of Science and Technology of China, Hefei 230026, People’s Republic of China; $^{19}$Chemistry Department, Humboldt-Universität zu Berlin, Germany; $^{20}$Department of Chemistry, University of North Texas, Denton, TX 76201, USA; $^{21}$Institute of Functional Interfaces, Karlsruhe Institute of Technology (KIT), 76344 Eggenstein-Leopoldshafen, Germany.}
\vspace{12pt}
} 

\date{\today}

\begin{abstract}
The expansive production of data in materials science, their widespread sharing and repurposing requires educated support and stewardship. 
In order to ensure that this need helps rather than hinders scientific work, the implementation of the FAIR-data principles ({\em Findable, Accessible, Interoperable, and Reusable}) must not be too narrow. Besides, the wider materials-science community ought to agree on the strategies to tackle the challenges that are specific to its data, both from computations and experiments. In this paper, we present the result of the discussions held at the workshop on ``Shared Metadata and Data Formats for Big-Data Driven Materials Science''. We start from an operative definition of metadata, and what features a FAIR-compliant metadata schema should have. We will mainly focus on computational materials-science data and propose a constructive approach for the {\em FAIRification} of the (meta)data related to ground-state and  excited-states calculations, potential-energy sampling, and generalized workflows. Finally, challenges with the {\em FAIRification} of experimental (meta)data and materials-science ontologies are presented together with an outlook of how to meet them.
\end{abstract}

\flushbottom
\maketitle
\thispagestyle{empty}


\section*{Introduction: Metadata and FAIR data principles}
\label{sec:intro}

The amount of data that has been produced in materials science till today and its day-by-day increase are massive\cite{rickman2019materials}. The dawn of the data-centric era\cite{HeyTansleyEtAl:2009} requires that such data are not just stored, but also carefully annotated in order to find, access, and possibly reuse them. Terms of good practice to be adopted by the scientific community for the management and stewardship of its data, the so-called FAIR-data principles, have been compiled by the FORCE11 group.\cite{wilkinson2016fair}
Here, the acronym FAIR stands for {\em Findable, Accessible, Interoperable, and Reusable}, 
which applies not only to {\em data}, but also to {\em metadata}. 
Other terms for the ``R'' in FAIR are ``repurposable'' and ``recyclable''. The former term indicates that data may be used for a different purpose than that for which they were initially created. The latter term hints at the fact that data in materials science are often exploited only once for supporting the thesis of a single publication and then they are stored and forgotten. In this sense, they would constitute a ``waste'' that can be recycled, provided that they can be found and they are properly annotated.

Before examining the meaning and importance of the four terms of the FAIR acronym, it is worth defining what metadata are with respect to data. To the purpose, we start by introducing the concept of {\em data object}. 
A {\em  data object} is the collective storage of information related to an elementary entry in a database. One can consider it as a row in a table, where the columns can be occupied by simple scalars, higher-order mathematical objects, strings of characters, or even full documents (or other media objects). 
In the materials-science context, a {\em  data object} is the collection of attributes (the columns in the above-mentioned table) that represent a material or, even more fundamentally, a {\em snapshot} of the material captured by a single configuration of atoms, or it may be a set of measurements from well-defined {\em equivalent samples} (\textcolor{black}{see below for a discussion on this concept)}.
For instance, in computational materials science, the attributes of a {\em  data object} could be both the inputs (e.g., the coordinates and chemical species of the atoms constituting the material, the description of the physical model used for calculating its properties), and the outputs (e.g., total energy, forces, electronic density of states, etc.) of a calculation. Logically and physically, inputs and outputs are at different levels, in the sense that the former determine the latter. Hence, one can consider the inputs as {\em metadata} describing the {\em data}, i.e., the outputs.  
In turn, the set of coordinates $A$ that are metadata to some observed quantities, may be considered as data that depend on another set of coordinates $B$, and the forces acting on the atoms in that set $A$. So, the set of coordinates $B$ and the acting forces are metadata to the set $A$, now regarded as data. Metadata can always be considered to be data as they could be objects of different, independent analyses than those performed on the calculated properties. In this respect, whether an attribute of a {\em  data object} is data or metadata depends on the {\em context}. This simple example also depicts a {\em provenance} relationship between the data and their metadata. 

The above discussion can be summarized in a more general definition of the term metadata: {\em Metadata are attributes that are necessary to locate, fully characterize, and ultimately reproduce other attributes that are identified as data.}
The metadata include a clear and unambiguous description of the data as well as their full provenance.
This definition is reminiscent of the definition given by NIST \cite{grassi2018attribute}: ``Structured information that describes, explains, locates, or otherwise makes it easier to retrieve, use, or manage an information resource. Metadata is often called data about information or information about information'' . With our definition, we highlight the role of data ``reproducibility'', which is crucial in science.

\textcolor{black}{Within the "full characterization" requirement, we highlight {\em interpretation} of the data as a crucial aspect. In other words, the metadata must provide enough information on a stored value (therein including, e.g., adimensional constants) to make it unambiguous whether two data objects may be compared with respect to the value of a given attribute or not.}

\textcolor{black}{Next, we should notice that, whereas in computational materials science the concept of data object identified by a single atomic configuration is well defined, in experimental materials science the concept of a class of {\em equivalent samples} is very hard to implement operationally. For instance, a single specimen can be altered by a measurement operation and thus cannot, strictly speaking, be measured twice. At the same time, two specimens prepared with the same synthesis recipe, may differ in substantial aspects due to the presence of different impurities or even crystal phases, thus yielding different values of a measured quantity. In this respect, here we use the term {\em equivalent sample} in its abstract, ideal meaning, but we also mention that one of the main purposes of introducing well-defined metadata in materials science is to provide enough characterization of experimental samples to put into practice the concept of {\em equivalent samples}.}

The need for storing and characterizing data by means of metadata is determined by two main aspects, related to data usage. The first aspect is as old as science: {\em reproducibility}. In an experiment or computation, all the necessary information needed to reproduce the measured/calculated data (i.e., the metadata) should be recorded, stored, and retrievable. The second aspect becomes prominent with the demand for {\em reusability}. Data can and should be also usable for purposes that were not anticipated at the time they were recorded.
A useful way of looking at metadata is that they are attributes of {\em  data objects} answering the ``wh- questions'': who, what, when, where, why, and how. For example, ``Who has produced the data?'', ``What are the data expected to represent (in physical terms)?'', ``When were they produced?'', ``Where are they stored?'', ``For what purpose were they produced?'', and ``By means of which methods were the data obtained?''. The latter two questions also refer to the concept of {\em provenance}, i.e., the logical sequence of operations that determine, ideally univocally, the data. Keeping track of the provenance requires the possibility to record the whole {\em workflow} that has lead to some calculated or measured properties (for more details, see Section ``Metadata for Computational Workflows''
). 

From a practical point of view, the metadata are organized in a schema. We summarise what the FAIR principles imply in terms of a metadata schema as follows:

\begin{itemize}
\item {\em Findability} is achieved by assigning unique and Persistent Identifiers (PIDs) to data and metadata, describing data with rich metadata, and {\em registering} (see below)
the (meta)data in searchable resources. Widely known examples of PIDs are digital object identifiers (DOIs) and (permanent) Uniform Resource Identifiers (URIs).
According to ISO/IEC 11179, a metadata {\em registry} (MDR) is a database of metadata that supports the functionality of registration. Registration accomplishes three main goals: identification, provenance, and monitoring quality. Furthermore, an MDR manages the semantics of the metadata, i.e., the relationships (connections) among them.

 \item {\em Accessibility} is enabled by ``application programming interfaces'' (APIs),  which allow one to query and retrieve single entries as well as entire archives.
 \item {\em Interoperability} implies the use of formal, accessible, shared, and broadly applicable languages for knowledge representation (these are known as formal ontologies and will be discussed in Section ``Outlook on ontologies in materials science''
 ), use of vocabularies to annotate data and metadata, and inclusion of references. 
 \item {\em Reusability}~
 hints at the fact that data in materials science are often exploited only once for a focus-oriented research project, and many data are not even properly stored as they turned out to be irrelevant for the focus. 
 In this sense, many data constitute a ``waste'' that can be recycled, provided that the data can be found and they are properly annotated. \\
\end{itemize}

Establishing one or more metadata schemas that are FAIR-data-principles compliant, and that therefore enable the materials-science community to efficiently share the heterogeneously and decentrally produced data, needs to be a community effort. The workshop ``Shared Metadata and Data Formats for Big-Data Driven Materials Science: A \mbox{NOMAD}--FAIR-DI Workshop'' was organized and held in Berlin in July 2019 to ignite this effort. In the following sections, we describe the identified challenges and first plans, divided into different aspects that are crucial to be addressed in computational materials science. We close with an outlook on a metadata schema for experimental materials science and on the introduction of formal ontologies for materials-science databases.

In the next Section, we describe the identified challenges and first plans for FAIR metadata schemas for computational materials science, where we also summarize as an example the main ideas behind the metadata schema implemented in the Novel-Materials Discovery (\mbox{NOMAD}) Laboratory for storing and managing millions of data objects produced by means of atomistic calculations (both {\em ab initio} and molecular mechanics), employing tens of different codes, which cover the overwhelming majority of what is actually used in terms of volume-of-data production in the community.
We then follow with more detailed sections discussing the specific challenges related to {\em interoperability} and {\em reusability} for ground-state calculations (Section ``Metadata for ground-state electronic-structure calculations'' 
), perturbative and excited-state calculations (Section ``Metadata for external-perturbation and excited-state electronic-structure calculations''
), potential-energy sampling (molecular-dynamics and more, Section ``Metadata for potential-energy sampling''
), and workflows in general (Section ``Metadata for Computational Workflows''
) are addressed in detail in the following sections. Challenges related to the choice of file formats are discussed in Section ``File Formats''
\textcolor{black}{An outlook on metadata schema(s) for experimental materials science and on the introduction of formal ontologies for materials-science databases constitute Sections ``Metadata schemas for experimental materials science''
and ``Outlook on ontologies in materials science''
, respectively.}

\section*{Towards FAIR metadata schemas for computational materials science}
\label{sec:metainfo}
The materials-science community has realized long ago that it is necessary to structure 
data by means of metadata schemas.
In this Section, we describe the pioneering and recent examples of such schemas, and how a metadata schema becomes FAIR-data-principles compliant.

\begin{figure*}[ht]
\centering
\includegraphics[width=0.75\textwidth]{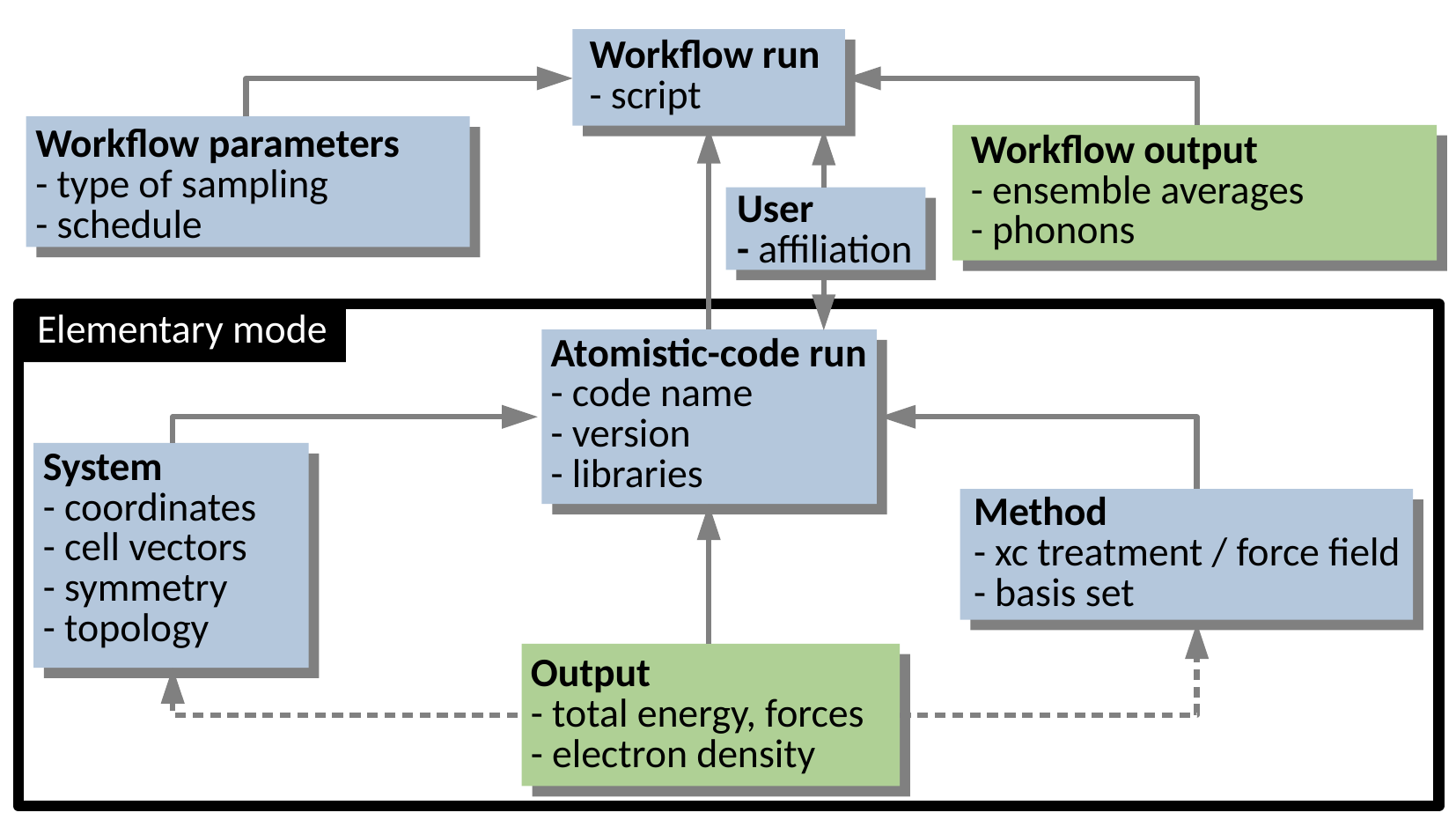}
\caption{Simplified schema of the {\em \mbox{NOMAD} Metainfo}. The rectangles symbolize the section-type metadata, for each section a few examples of therein contained quantity-type or (sub)sections metadata are listed. Sections are always written in bold font. The solid arrows stand for the {\em is contained in} relationship, while the dashed arrows are for the {\em has reference in} relationship.}
\label{fig:flowchart}
\end{figure*}

\textcolor{black}{To our knowledge, the first systematic effort to build a metadata schema for exchanging data in chemistry and materials science is CIF, an acronym that originally stood for Crystallographic Information File, the data-exchange standard file format introduced in 1991 by Hall, Allen and Brown \cite{hall1991crystallographic,bernstein2016specification}. Later, the CIF acronym was extended to also mean Crystallographic Information Framework \cite{hall2005crystallographic}, a broader system of exchange protocols based on data dictionaries and relational rules expressible in different machine-readable manifestations. These include the Crystallographic Information File itself, but also, for instance, XML (eXtensible Markup Language), a general framework for encoding text documents in a format that is meant to be at the same time human and machine readable. CIF was developed by the International Union of Crystallography (IUCr) working party on Crystallographic Information and was adopted in 1990 as a standard file structure for the archiving and distribution of crystallographic information. It is now well established and is in regular use for reporting crystal structure determinations to Acta Crystallographica and other journals. More recently, CIF has been adapted to different areas of science such as structural biology (mmCIF, the macromolecular CIF \cite{westbrook20055}) and spectroscopy \cite{el2019raman}. The CIF framework includes strict syntax definition in a machine readable form and dictionary defining (meta)data items. It has been noted that the adoption of the CIF framework in IUCr publications has allowed for a significant reduction of the amount of errors in published crystal structures \cite{mcmahon1996role,brown2002cif}
}

An early example of an exhaustive metadata schema for chemistry and materials science is the Chemical Markup Language (CML\cite{CML3,CML1,CML2}), whose first public version was released in 1995. CML is a dictionary, encoded in XML for chemical metadata.  CML is accessible (for reading, writing, and validation) via the Java library JUMBO (Java Universal Molecular/Markup Browser for Objects \cite{CML2}).  The general idea of CML is to represent with a common language all kinds of documents that contain chemical data, even though currently the language --- as of the latest update in 2012\cite{b_note_CML_schema} --- covers mainly the description of molecules (e.g., IUPAC name, atomic coordinates, bond distances) and of inputs/outputs of computational chemistry codes such as Gaussian03 \cite{g03} and NWChem \cite{NWChem}. Specifically, in the CML representation of computational chemistry calculations \cite{cml_example}, (ideally) all the information on a simulation that is contained in the input and output files is mapped onto a format that is in principle independent of the code itself. Such information is: 
\begin{itemize}
 \item Administrative data like the code version, libraries for the compilation, hardware, user submitting the job;
 \item Materials-specific (or materials-snapshot-specific) data like computed structure (e.g., atomic species, coordinates), the physical method (e.g., electronic exchange-correlation treatment, relativistic treatment), numerical settings (basis set, integration grids, etc.);
 \item Computed quantities (energies, forces, sequence of atomic positions in case a structure relaxation or some dynamical propagation of the system is performed, etc \ldots).\\
\end{itemize}

The different types of information are hierarchically organized in {\em modules}, e.g., {\em environment} (for the code version, hardware, run date, etc.), {\em initialization} (for the exchange correlation treatment, spin, charge), {\em molgeom} (for the atomic coordinates and the localized basis set specification), {\em finalization} (for the energies, forces, etc.). The most recent release of the CML schema contains more than 500 metadata-schema items, i.e., unique entries in the metadata schema. \textcolor{black}{It is worth noticing that CIF is the dictionary of choice for the crystallography domain within CML.}
Another long-standing activity is JCAMP-DX (Joint Committee on Atomic and Molecular Physical Data - Data Exchange)\cite{mcdonald1988jcamp}, a standard file format for exchange of infrared spectra and related chemical and physical information was established in 1988 and then updated with IUPAC recommendations until 2004. It contains standard dictionaries for infrared spectroscopy, chemical structure, NMR\cite{davies1993jcamp}, and mass \cite{lampen1994jcamp} and ion-mobility spectrometry \cite{baumbach2001jcamp}.
The European Theoretical Spectroscopy Facility (ETSF) File Format Specifications were proposed in 2007 \cite{gonze2007extensible,gonze2008specification,caliste2008sharing}, in the context of the European Network of Excellence NANOQUANTA, in order to overcome widely known portability issues of input/output file formats across platforms. 
The Electronic Structure Common Data Format (ESCDF) Specifications\cite{ghiringhelli2017towards} is the ongoing continuation of the ETSF project and is part of the CECAM Electronic Structure Library, a community-maintained collection of software libraries and data standards for electronic-structure calculations \cite{Oliveira2020}.

The largest databases of computational materials-science data, AFLOW\cite{aflowlib}, Materials Cloud \cite{MaterialsCloud}, Materials Project\cite{Jain2013}, the \mbox{NOMAD} Repository and Archive \cite{draxl2018nomad,draxl2019nomad,draxl2020big}, OQMD\cite{kirklin2015open}, and TCOD\cite{merkys2017posteriori} offer application programming interfaces (APIs) that rely on dedicated metadata schemas. Similarly, AiiDa\cite{AiiDA-Pizzi2016,huber2020aiida,uhrin2021workflows} and ASE\cite{larsen2017atomic}, which are schedulers and workflow managers for computational materials-science calculations, adopt their own  metadata schema.
OpenKIM\cite{tadmor2011potential} is a library of interatomic models (force fields) and simulation codes that test the predictions of these models, complemented with the necessary first-principles and experimental reference data. Within OpenKIM, a metadata schema is defined for the annotation of the models and reference data.
Some of the metadata in all these schemas are straightforward to map onto each other (e.g., those related to the structure of the studied system, i.e., atomic coordinates and species, and simulation-cell specification), others can be mapped with some care. 
The OPTIMADE (Open Databases Integration for Materials Design\cite{Andersen2021}) consortium has recognized this potential and has recently released the first version of an API that allows users to access a common subset of metadata-schema items, independent of the schema adopted for any specific database/repository that is part of the consortium.

In order to clarify how a metadata schema can explicitly be FAIR-data-principles compliant, we describe as an example the main features of the {\em \mbox{NOMAD} Metainfo}, onto which the information contained in the input and output files of atomistic codes, both {\em ab initio} and force-field based, is mapped. The first released version of the {\em \mbox{NOMAD} Metainfo} is described in Ref. \cite{ghiringhelli2017towards} and it has powered the NOMAD Archive since the latter went online in 2014, thus predating the formal introduction of the FAIR-data principles \cite{wilkinson2016fair}. 

Here, we give a simplified description, graphically aided by Fig. \ref{fig:flowchart}, which highlights the hierarchical/modular architecture of the metadata schema. 
The {\em elementary mode} in which an atomistic materials-science code is run  
(encompassed by the black rectangle) serves the computation of some observables (\emph{Output}) for a given \emph{System}, specified in terms of atomic species arranged by their coordinates in a box, and for a given physical model (\emph{Method}), including specification of its numerical implementation.  
Sequences or collections of such runs are often defined via a {\em Workflow}.
Examples of workflows are:
\begin{itemize}
    \item Perturbative physical models (e.g., second-order Møller–Plesset, MP2, Green's function based methods such as $G_0W_0$, random-phase approximation, RPA) evaluated using self-consistent solutions provided by other models (e.g., density-functional theory, DFT, Hartree-Fock method, HF) applied on the same {\em System};
    \item Sampling of some desired thermodynamic ensemble by means of, e.g., molecular dynamics;
    \item Global- and local-minima structure searches; 
    \item Numerical evaluations of equations of state, phonons, or elastic constants by evaluating energies, forces, and possibly other observables;
    \item Scans over the compositional space for a given class of materials (high-throughput screening). \\
\end{itemize}
%
\textcolor{black}{The workflows can also be nested, e.g., a scan over materials (different compositions and/or crystal structures), contains a local optimization for each material and extra calculations based on each local optimum structure such as evaluation of phonons, bulk modulus, or elastic constants, etc.}

The {\em \mbox{NOMAD} Metainfo} organizes metadata into sections, which are represented in Fig. \ref{fig:flowchart} by the labeled boxes. The sections are a {\em type} of metadata, which group other metadata, e.g., other sections or {\em quantity}-type metadata. The latter are metadata related to scalars, tensors, strings, which represent the physical quantities resulting form calculations or measurements. In a relational-database model, the sections would correspond to tables, where the {\em  data objects} would be the rows, and the quantity-type metadata the columns. 
In its most simple realization, a metadata schema is a {\em key-value} dictionary, where the key is a name identifying a given metadata. In {\em NOMAD Metainfo}, similarly to CML, the key is a complex entity grouping the several attributes.
Each item in {\em NOMAD Metainfo} has {\em attributes}, starting with its {\em name}, a string that must be globally unique, well-defined, intuitive, and as short as possible. Other attributes are the human-understandable {\em description}, which clarifies the meaning of the metadata, the {\em parent section}, i.e., the section the metadata belongs to, and the {\em type}, whether the metadata is, e.g., a section or a {\em quantity}.  Another possible {\em type}, the {\em category} type, will be discussed below. For the quantity-type metadata, other important attributes are {\em physical units} and {\em shape}, i.e., the dimensions (scalar, vector of a certain length, a matrix with a certain number of rows and columns, etc.), and {\em allowed values}, for metadata that admit only a discrete and finite set of values. 

All definitions in the {\em \mbox{NOMAD} Metainfo} have the following attributes:
\begin{itemize}
\item A globally unique qualified name; 
\item Human-readable/interpretable description and expected format (e.g., scalar, string of a given length, array of given size);
 \item Allowed values; 
 \item Provenance, which is realized in terms of a hierarchical and modular schema, where each {\em  data object} is linked to all the metadata that concur to its definition. Related to provenance, an important aspect of {\em \mbox{NOMAD} Metainfo} is its {\em extensibility}. It stems from the recognition that reproducibility is an empirical concept, thus at any time, new, previously unknown or disregarded metadata may be recognized as necessary. The metadata schema must be ready to accommodate such extensions seamlessly.\\
\end{itemize}

The representation in Fig. \ref{fig:flowchart} is very simplified for tutorial purposes. For instance, a workflow can be arbitrarily complex. In particular, it may contain a hierarchy of sub-workflows. 
In the currently released version of the {\em \mbox{NOMAD} Metainfo}, the elementary-code-run modality is fully supported, i.e., ideally all the information contained in a code run is mapped onto the metadata schema. However, the workflow modality is still under development.
An important implication of the hierarchical schema is the mapping of  any (complex) workflow onto the schema. That way, all the information obtained by its steps is stored. This is achieved by parsers, which have been written by the \mbox{NOMAD} team for each supported simulation code. One of the outcomes of the parsing is the assignment of a PID to each parsed {\em  data object}, thus allowing for its localization, e.g., via a URI. 

\begin{figure}[h!]
\centering
\includegraphics[width=\columnwidth]{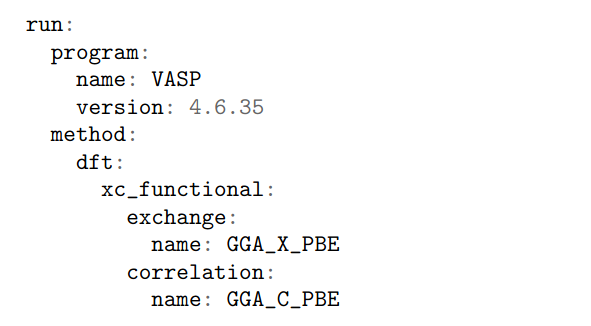}
\caption{A portion of a YAML file instantiating Metainfo for one entry of the \mbox{NOMAD} Archive.}
\label{fig:snippet}
\end{figure}

The {\em \mbox{NOMAD} Metainfo} is inspired by the CML, in particular in being hierarchical/modular. 
\textcolor{black}{Each instance of a metadata-schema is uniquely identified, so that it can be associated with a URI for its convenient accessibility. An instance of a metadata-schema can be generated by using a dedicated parser by pairing each parsed value with its corresponding metadata label. As an example, below (Fig. \ref{fig:snippet} we show a portion of the YAML file (see section ``File Formats''
) instantiating Metainfo for a specific entry of the \mbox{NOMAD} Archive.} 

\noindent This entry can be searched by typing:
\begin{center}
\texttt{entry\_id=zvUhEDeW43JQjEHOdvmy8pRu-GEq} 
\end{center}
in the search bar at \url{https://nomad-lab.eu/prod/v1/gui/search/entries}. In this example, key-value pairs are visible as well as the nested-section structuring.

The modularity and uniqueness together allow for a straightforward extensibility including customization, i.e., introduction of metadata-schema items that do not need to be shared among all users, but may be used by a smaller subset of users, without conflicts. 

In Fig. \ref{fig:flowchart}, the solid arrows stand for the relationship {\em is contained in} between section-type metadata. A few examples of quantity-type metadata are listed in each box/section. Such metadata are also in an {\em is-contained-in} relationship with the section they are listed in. The dashed arrows symbolize the relationship {\em has reference in}. In practice, in the example of an {\em Output} section, the quantity-type metadata contained in such a section are evaluated for a given system described in a {\em System} section and for a given physical model described in a {\em Method} section. So, the section {\em Output} contains a reference to the specific {\em System} and {\em Method} sections holding the necessary input information. At the same time, the {\em Output} section {\em is contained in} a given {\em Atomistic-code run} section. These relationships among metadata already build a basic ontology, induced by the way computational data are produced in practice, by means of workflows and code runs. This aspect will be reexamined in Section ``Outlook on ontologies in materials science''.

We now come to the {\em category-type} metadata that allow for complementary, arbitrarily complex ontologies to be built by starting from the same metadata. They define a concept, such as ``energy'' or ``energy component'', in order to specify that a given quantity-type metadata has a certain meaning, be it physical (such as ``energy'') or computer-hardware related, or administrative.  To the purpose, each section and quantity-type metadata is related to a category-type metadata, by means of an {\em is-a} kind of relationship. Each category-type metadata can be related to another category-type metadata by means of the same {\em is-a} relationship, thus building another ontology on the metadata, which can be connected with top-down ontologies such as EMMO\cite{cemmo2020} (see section ``Outlook on ontologies in materials science''
for a short description of EMMO).

The current version of {\em \mbox{NOMAD} Metainfo} includes more than 400 metadata-schema items. More specifically, these are the {\em common} metadata, i.e., those that are code-independent. Hundreds more metadata are code-specific, i.e., mapping pieces of information in the codes' input/output that are specific to a given code and not transferable to other codes. The {\em \mbox{NOMAD} Metainfo} can be browsed at \url{https://nomad-lab.eu/prod/v1/gui/analyze/metainfo}.

To summarize, the {\em \mbox{NOMAD} Metainfo} addresses the FAIR-data principles in the following sense:
\begin{itemize}
    \item {\em Findability} is enabled by unique names and a human-understandable description; 
    \item {\em Accessibility} is enabled by the PID assigned to each metadata-schema item, which can be accessed via a RESTful \cite{fielding2000architectural} API (i.e., an API supporting the access via web services, through common protocols, such as HTTP), specifically developed for the {\em \mbox{NOMAD} Metainfo}.
    Essentially all \mbox{NOMAD} data are open access and users who wish to search and download data do not need to identify themselves. They only need to accept the CC BY license. Uploaders can decide for an embargo. These data are then shared with a selected group of colleagues. 
    \item {\em Interoperability} is enabled by the extensibility of the schema and the category-type metadata, which can be linked to existing and future ontologies (see Section ``Outlook on ontologies in materials science'').
    \item {\em Reusability/Repurposability/Recyclability} is enabled by the modular/hierarchical structure that allows for accessing calculations at different abstraction scales, from the single observables in a code run to a whole complex workflow (see Section ``Metadata for Computational Workflows'').\\
\end{itemize}

The usefulness and versatility of a metadata schema are demonstrated by the multiple access modalities it allows for. The {\em \mbox{NOMAD} Metainfo} schema is the basis of the whole \mbox{NOMAD} Laboratory infrastructure, which supports access to all the data in the \mbox{NOMAD} Archive, via the \mbox{NOMAD} API (also an implementation of the OPTIMADE API\cite{Andersen2021} is supported). This API powers three different access modes of the Archive: the {\em Browser}\cite{browser}, which allows searches for single or groups of calculations, the {\em Encyclopedia}\cite{ency}, which display the content of the Archive organized by {\em materials}, and the {\em Artificial-Intelligence (AI) Toolkit}\cite{ghiringhelli2021ai,sbailo2022nomad,ait}, which connects in Jupyter notebooks script-based queries and AI (machine-learning, data-mining) analyses of the filtered data. All the three services are accessible via a web browser running the dedicated GUI offered by \mbox{NOMAD}. 

\section*{Metadata for ground-state electronic-structure calculations}
\label{sec:gs}

By ground-state calculations, we mean calculations of the electronic structure --- e.g., \textcolor{black}{eigenvalues and eigenfunctions of the single-particle Kohn-Sham equations}, the electron density, the total energy and possibly its derivatives (forces, force constants) --- for a fixed configuration of nuclei. This refers to a point located on the Born-Oppenheimer potential-energy surface, and is a necessary step in geometry optimization, molecular dynamics, the computation of vibrational (phonon) spectra or elastic constants, and more. Thus, ground-state calculations represent the most common task in computational materials science, and the involved approximations are relatively well established. For this reason, they are already extensively covered by the {\em \mbox{NOMAD} Metainfo}. Nevertheless, some challenges in defining metadata for such calculations still remain, as discussed below. In particular, density-functional theory (DFT) is the workhorse approach for the great majority of ground-state calculations in materials science. 
Highly accurate quantum-chemistry models are more computationally expensive than DFT and their use in applications is less widespread. However, they can provide accurate benchmark references for DFT, making high-quality quantum-chemical data essential also for DFT-based studies. Below we analyze the ground-state electronic structure calculations mainly in reference to DFT, but most of the stated principles are also valid for quantum-chemical calculations. A detailed discussion of the latter is deferred to Section ``Quantum-chemistry methods''.

\subsection*{Approximations to the DFT exchange-correlation functional}
Approximations to the DFT exchange-correlation (xc) functionals are identified by a name or acronym (e.g., “PBE”), although sometimes this identification is not unique or complete. 
As metadata, we suggest to use the identifiers of the 
Libxc library~\cite{Marques2012_2272,Lehtola2018_1},
which is the largest bibliography of xc functionals. 
In order to be both human and computer friendly, the Libxc identifiers consist of a human-readable string that has a unique integer associated with it.
Often, the above-noted identification needs some refinement, because xc functionals typically depend on a set of parameters and these may be modified for a given calculation. 
Obviously, there is a need to standardize the way in which such parameters are referenced.
Just like it is possible to use the Libxc identifiers for the functionals themselves, one may also use the Libxc naming scheme for their internal parameters. Obviously, code developers have to ensure that this information is contained in the respective input and/or output files. As Libxc provides version numbers of the xc functionals, it is important that this information is also available.

\subsection*{Basis sets}
Complete and unambiguous specification of the basis set is crucial for judging the precision of a calculation.
Ground-state calculations should include the full information about the basis sets used, including a DOI that a basis may be referred to. The use of repositories of basis sets, like the Basis Set Exchange repository~\cite{pritchard2019new}, is therefore strongly recommended.

Basis sets can be coarsely divided into two classes, i.e., atom-position-dependent (atom-centered, bond-centered) and cell-dependent (such as plane waves) ones. Also 
a combination of both is possible, as, e.g., realized in augmented plane-wave or projector-augmented wave methods.
For the atom-centered basis, the list of centers needs to be provided, and these may even contain positions where no actual atomic nucleus is located. 
The {\em \mbox{NOMAD} Metainfo} contains a rather complete set of metadata to describe atom-centered basis sets.
A more complete description of cell-dependent basis sets can be found in the ESCDF, which is planned to be merged with the {\em \mbox{NOMAD} Metainfo}. 

\subsection*{Energy reference}
In order to enable {\em interoperability} and {\em reusability} of energies computed with different electronic-structure methods, it is necessary to define a ``general energy zero''. An analysis of this problem and some clues on how to tackle it were already discussed by some of us in a previous work\cite{ghiringhelli2017towards}. The following is a further attempt to advance and systematize ideas and solutions.

The problem of comparing energies is not restricted to computational materials science and chemistry. In fact, it also arises in experimental chemistry, as for instance, only enthalpy or entropy differences can be measured, but not absolute values. To solve this, chemists have defined a reference state for each element, called the {\em standard state}, which is defined as the element in its natural form at standard conditions, while the {\em heat of formation} is used to measure the change from the elements to the compound. In computational materials science and chemistry, we can adopt a similar approach. For each element we need to define a reference system as the zero of the energy scale. To do so, we introduce some definitions:

\begin{itemize}
 \item A system is a defined set of one or more atoms, with a given geometry and, if periodic, a given unit cell. It can be an atom, a molecule, a periodic crystal, etc. If relevant, the charge, the spin-state or magnetic ordering needs to be specified. 
 \item A reference system is a well-defined system to which other systems are compared to. 
 \item A calculated energy is the energy obtained by a numerical simulation of a system with given input data and parameters, defining the Hamiltonian (i.e., DFT xc-functional approximation) or the many-electron model (e.g., Hartree-Fock, MP2, ``coupled-cluster singles, doubles, and perturbative triples'', CCSD(T)), the basis set, and the numerical parameters.\\
\end{itemize}
 Whether the reference system is an atom, an element in its natural form, some molecule or other system, does not matter, as long as it is well defined. Defining the system by atoms requires specifying how the orbitals are occupied, whether the atom is spherical, spin-polarized, etc. 
For each computational method and numerical settings, the energy per atom of the reference system must be calculated. The {\em standard energy} is then obtained by subtracting these values (multiplied by the number of constituents) from the calculated total energy. 
For example, to determine the energy of formation of a molecule like H$_2$O or a crystal like SiC, we calculate the difference in total energies $E(\mathrm{H}_2\mathrm{O}) - E(\mathrm{H}_2) - \frac{1}{2} E (\mathrm{O}_2)$, or $E(\mathrm{SiC}) - E(\mathrm{Si}) - E(\mathrm{C})$, respectively. Here, $\mathrm{H}_2$ and $\mathrm{O}_2$ are isolated, neutral molecules while Si and C are free, neutral atoms. However, using the energy per atom of Si and C in their crystalline ground-state structure would be an option as well.   
We propose to tabulate the reference energies for the most common computational methods, so that they can be applied without further computations and preferably automatically by the codes themselves. 

Finally, we need to define what is meant by a computational method. The Hamiltonian and DFT functional are clearly part of the definition as is the basis set and the potential shape (including pseudopotentials (PP) and effective core potentials). The specific implementation may also be relevant.  
Gaussian-based molecular-orbital codes may give the same energy for an identical setup (see Section ``Quantum-chemistry methods''
), while plane-wave DFT codes may not.

One factor here is the choice of the PP. Irrespective of the used method, the computational settings determine the quality of a calculation. Most decisive here is the basis-set cut-off. 
For the plane-wave basis, convergence with respect to this parameter is straightforward. 
In any case, depending on the code, the method and details of the calculation, care needs to be taken to define all the adjustable parameters that significantly affect the energy when defining computational methods.

To tabulate standard energies, as suggested above, every computational method needs to be applied to all reference systems. This requires care in choosing the reference systems to ensure that an as-wide-as-possible range of codes and methods are actually suited for these calculations. It may be that some codes cannot constrain the occupancies of atoms, or keep them spherical, which would be a problem if spherical atoms were chosen as the reference. Clearly, periodic crystals such as silicon are not suitable for molecular codes. 
It is possible, however, that some other codes could help bridging this gap. For example, FHI-aims\cite{blum2009ab} is not only capable of simulating crystalline system, but can also handle atoms and molecules and it can employ Gaussian-type orbitals (GTO) basis sets.
Thus, FHI-aims is able to reproduce energy differences between atoms/molecules and crystals. In this way, it can support codes such as Gaussian\cite{g03} or GAMESS\cite{GAMESS}. 

\section*{Metadata for external-perturbation and excited-state electronic-structure calculations}
\label{sec:xs}

A direct link from the DFT ground state (GS) to excitations is provided by time-dependent DFT (TDDFT). Alternatively, charged and neutral electronic excitations are described by means of Green-function approaches from many-body perturbation theory (MBPT). This route is predominantly (but not exclusively) used for the solid state, while TDDFT and quantum-chemistry approaches are typically preferred for finite systems. For strongly correlated materials, in turn, dynamical mean-field theory (DMFT) is often the methodology of choice, potentially combined with DFT and Green-function methods. Lattice excitations, if not directly treated by DFT molecular dynamics, are often handled by density-functional perturbation theory (DFTP); for their interaction with light, also Green-function techniques are used. DFPT not only allows for the description of vibrational properties, but also for treating macroscopic electric fields, applied macroscopic strains, or combinations of these. The type of perturbation is intimately related to the physical properties of interest, e.g., harmonic and anharmonic phonons, effective charges, Raman tensors, dielectric constants, hyper-polarizabilities, and many others. 

Characterizing the corresponding research data is a very complex and complicated task, for various reasons. First, such calculations rely on an underlying ground-state calculation, and thus carry along all uncertainties from it. Second, the methodology for excited states is scientifically and technically more involved by including many-body effects that govern diverse interactions. The methods thus rely on various, often not fully characterized approximations.

\subsection*{Diagrammatic techniques and TDDFT}
\label{sec:mbpt}
The most common application of $GW$ is to compute quasi-particle energies, i.e., energies that describe the removal or addition of a single electron. For this, the many-body electron-electron interaction is described by a two-particle operator, called the electronic self-energy. To compute this object, on the technical side we may need an additional (auxiliary) basis set, not the same as the one used in the ground-state calculation, coming with additional parameters. Likewise, there are various ways for doing the analytical continuation of the Green's function, as there are various ways for carrying out the required frequency integration, possibly employing a plasmon-pole model as an approximation. And there are also different ways of how to evaluate the screened Coulomb potential $W$. Most important is the flavor of $GW$, i.e., whether it is done in a single-shot manner, called $G_0W_0$, or in a self-consistent way. If the latter, what kind of self-consistency (scf) is used –-- any type of partial scf, quasi-particle scf, or any other type which would remedy any starting-point dependence, i.e., the dependence of the results on the xc functional of the initial DFT (or Hartree-Fock or alike) used in the GS. 

While $GW$ is the method of choice for quasi-particle energies (and potentially also life times) within the realm of MBPT, we need to solve the Bethe-Salpeter equation (BSE) to tackle electron-hole interactions. This approach should typically be applied on top of a $GW$ calculation, but often the quasi-particle states are approximated by DFT results adjusted by a scissors operator to widen the band gap in a similar way to the latter. In all cases, BSE carries along all subtleties from the underlying steps. In addition, it comes with its own issues, like the way of screening the Coulomb interaction (electron-hole this time), the representation of non-local operators, and alike.

DMFT, as a rather young and quickly developing field, naturally experiences a plenitude of “experimental” implementations, differing in many aspects, with one of the major obstacles being the quite vast amount of combinations of software. Some of the approaches are computationally light, allowing for the construction of model Hamiltonians based on DFT calculations; others are computationally too demanding and can be applied only to simple systems with a few orbitals; most of the methods rely on Green’s functions and self-energies. Diagrammatic extensions beyond standard DMFT methods employ various kinds of vertex functions. Other issues concern the definition of how to handle the Coulomb interactions, where the parameters can either be chosen empirically or can be calculated by first principles.

Specific issues of TDDFT concern, in a first place, the distinction between the linear-response regime and the time-propagation of the electronic states in presence of a time-dependent potential. For the former, the xc kernel plays the same role as the xc functional of the GS, raising (besides numerical precision) questions related to accuracy. For the latter, there are various ways and flavors for how to implement the time-evolution operator. Moreover, one can write this operator as a simple exponential or use more elaborate expressions, like the Magnus expansion or the enforced time reversal symmetry. Regarding the exponential, one can employ a Crank-Nicolson expansion, expand in a Taylor series or employ Houston states. Obviously, each of them comes with approximations and additionally, numerical issues.

In summary, all the variety captured by the different methods together with the related multitude of computational parameters, needs to be carefully reflected by the metadata schema. This is not only imperative for ensuring reproducible results but also for evaluating the accuracy of methods and commonly used approximations. Besides, further subtleties related to algorithms in the actual implementations in different codes requires the code developers to embark on this challenge. 

\subsection*{Density-functional perturbation theory}
\label{DFPT}
Density-functional perturbation theory is used to obtain physical properties that are related to the (density-)response of the system to external perturbations, like the displacement potential according to lattice vibrations. Also in this case, the calculation relies on a preliminary GS run, inheriting all issues therefrom. After having chosen the type of perturbation, which requires method-dependent definitions and inputs, one needs to choose the order of perturbation: The linear response approach, that is implemented in many codes \textcolor{black}{(e.g.,  VASP \cite{kresse1996efficient}, octopus \cite{marques2003octopus}, CASTEP \cite{segall2002first}, FHI-aims\cite{blum2009ab}, Quantum Espresso \cite{giannozzi2009quantum}, ABINIT \cite{gonze2009abinit})}, allows for the determination of second-order derivatives of the total energy. Among these codes, some of them also allow for the calculation of third-order derivatives, like anharmonic vibrational effects. The variation of the Kohn-Sham orbitals can be obtained from the Sternheimer equation, where different methods are used for deriving its solution (iterative methods, direct linearization, integral formulation). 

\subsection*{Quantum-chemistry methods}
\label{sec:qc}
Quantum chemistry offers several methodological hierarchies for calculating quantities related to excited states, such as excitation energies, transition moments, ionization potentials, etc. 
As high-quality methods are computationally intensive, without additional approximations such methods can be applied to relatively small molecular systems only. 

Among the standard quantum chemical approaches that can be routinely applied to study excited states of small to medium-sized molecules one can distinguish two large groups, i.e., single-reference and multi-reference methods. The single-reference coupled-cluster (CC) hierarchy for excited states can be formulated in terms of the so-called equation-of-motion approach or time-dependent linear response. 

Generally, for well-behaving closed-shell molecules, the single-reference quantum-chemical methods can be used as a black box. The formalisms of the MP$n$ and CC models are uniquely defined and well documented. The GTO basis sets from the standard basis set families (Pople, Dunning, etc.) are also uniquely defined by the acronym. In practical implementations of these methods, of course various thresholds are usually introduced for prescreening, convergence, etc., but the default values for these thresholds are routinely set very conservatively to guarantee a sub-microhartree precision of the final total energies. Problems might, however, arise due to the iterative character of most of the mentioned techniques, as convergence to a certain state (both in the ground-state and/or excited-state parts of the calculation) depends on starting guess, preconditioner, possible level shifts, type of convergence accelerator, etc. Unfortunately, the parameters that control the convergence are often not sufficiently well documented and might not be found in the output. Such problems mainly occur in open-shell cases (note that in the Delta methods at least one of the calculations has to involve an open-shell system). Sometimes a cross-check between several codes becomes essential to detect convergence faults.
 
When it comes to larger systems and approximate CC models are utilized, the importance of the involved tolerances and underlying protocols substantially increases. The approximations can include, for example, the density-fitting technique, local approximation, Laplace transform, and others. Important parameters here are the auxiliary basis set, the fitting metric, the type of fitting (local or non-local), and if local, how the fit-domains are determined, etc. The result of the calculations that use local correlation techniques are influenced by the choice of the virtual space and the corresponding truncation protocols and tolerances, the pair hierarchies and the corresponding approximations for the CC terms, etc. For Laplace-transform-based methods, the details of the numerical quadrature matter. Unfortunately, these subtleties are very specific and technical and even if given in the output, can hardly be properly understood and analyzed by non-specialists who are not involved in the development of the related methods. Therefore, the protocols behind the approximations are usually appropriately automatized, and the defaults are chosen such that for certain (benchmarking) sets of systems the deviations in the energy are substantially smaller than the expected error of the method itself (e.g., 0.01 eV for the excitation energy). However, for these methods, additional benchmarks and cross-checks between different programs and approaches would be very important. 

Multi-reference methods come with quite a number of different flavors, where the most widely used ones are complete active-space self-consistent Field (CASSCF), complete active-space second-order perturbation theory (CASPT2), and multireference configuration interaction (MRCI). 
For difficult cases (e.g., strongly correlated systems), these methods might remain the only option to obtain qualitatively and quantitatively correct result. Unfortunately, compared to the single reference methods, they are computationally expensive and much less of a black box. First of all, for each calculation one has to specify the active space or active spaces. The results may depend dramatically on this choice. Furthermore, the underlying theory is not always uniquely defined by the used acronym. For example, different formulations of CASPT2, MRCI, or other theories are not mutually equivalent depending on whether and how much internal contraction is used and additional approximations that neglect certain terms (e.g., many-electron density matrices) can be implicitly invoked. Besides, certain deficiencies of these methods, such as for example lack of size consistency in MRCI or intruder states in CASPT2, are often corrected by additional (sometimes empirical) schemes, which again are not always fully specified. All this makes the interpretation of deviations in results and cross-checks of these methods less conclusive. 

To summarize, quantum-chemical methods offer an excellent toolbox for accurate {\it ab initio} calculations for molecules (especially so for small and medium sized ones). However, severe issues concerning reproducibility and replicability remain, in particular for extended and/or open-shell systems. This calls for a more detailed specification of the implemented techniques by the developers, for example, a better design of the outputs, and a thorough analysis and documentation of the employed methods and parameters by the users. \textcolor{black}{A possible strategy addressing these issues would be two-fold.
a) Promoting the compliance of the developed software with the FAIR principles for software \cite{lamprecht2020towards,barkerIntroducingFAIRPrinciples2022}, which comprise the recommendation to publish the software in a repository with version control, have a well-defined license, register the code in a community registry, assign to each version a PID, and enable its proper citation \cite{katz2020recognizing,smith2016software}. Reproducibility can be enhanced by publishing software code under the Free/Libre Open-Source Software (F/LOSS) \cite{FOSS,FLOSS}  license and by documenting the computation environment (hardware, operating system version, computational framework and libraries that were used, if any)
b) Creation of well-defined benchmark datasets. Interoperability among different implementations of (in the intention) the same theoretical model can be assessed by the quantitative comparison over different codes (including different versions thereof) of a set of properties on an agreed-upon set of materials. Such datasets would obviously need to be stored in a FAIR-data-compliant fashion. A large community-based effort in this direction is being carried on in the DFT community \cite{lejaeghere2016reproducibility}, while in the many-body-theory community, implementation of this idea is just at its beginning\cite{schafer2021tracking})
}

\section*{Metadata for potential-energy sampling}
\label{sec:md}

Molecular dynamics (MD) simulations --- i.e., following the time evolution of a system, employ either \emph{ab initio} calculated forces and energies (aiMD) or molecular mechanics (MM) i.e., forces and energies are defined through empirical atomistic and coarse-grained potentials. The FAIR storing and sharing of their inputs and outputs comes with a number of specific challenges in comparison to single-point electronic-structure calculations.

Conceptually, aiMD and MM are similar, as a sequence of system configurations is evolved at discrete time steps. Positions, velocities, and forces at a given time step are used to evaluate positions and velocities, and hence forces in the new configuration, and so on.
In practice, MM simulations are orders of magnitude faster than aiMD, enabling much longer time scales and/or much larger system sizes.
Even though the trend towards massive parallelization will enable aiMD in the near future system to handle sizes comparable to today's standards for MM simulations, the latter will probably always enable larger systems. However, with machine-learned potentials and active learning techniques for their training, aiMD and MM may grow together in the future.  

In this Section, we focus on challenges more specific to MM simulations, having in mind large length scales, long time scales, and complex phase-space-exploration algorithms and workflows. They can be summarized as follows: 

\begin{enumerate}[label=(\roman*)]
\item In many cases, the investigated systems feature thousands of atoms with complex short- and long-range order and disorder, e.g., describing microstructural evolution such as crack propagation. This requires large, complex simulation cells with a range of chemical species to be correctly described and categorized.
\item 
Force-fields exist in a wide variety of flavors that require proper classification. On top of that, they allow for granular fine-tuning of the interactions, even for individual atoms. Faithfully representing complex force fields thus requires to also capture the chemical-bonding topology that is often needed to define the actual interactions.
\item 
The large length and long time scales presently come together with a multitude of simulation protocols, which use specific boundary conditions, thermostats, constraints, integrators, etc. The various approaches enable the computations of additional observables to be computed as statistical averages or correlations. 
Representing these properties implies the need to efficiently store and access large volumes of data, e.g.,  trajectories, including positions, and possibly also velocities and forces, for each atom at each time step.
\end{enumerate}

For the purpose of illustration, we start by identifying some typical use cases, then describe what is currently implemented in the \mbox{NOMAD} infrastructure and what is missing. The examples we adopt fall into two classes: (i) high throughput systems that are individually \emph{simple} ($\sim1\,000 - 10\,000$ particles) where the value of sharing comes from the ability to run analysis across many variants of, e.g., chemical composition or force field; (ii) sporadic simulations of very large systems or very long time scales which cannot readily be repeated by other researchers and thus are individually valuable to share.
Examples of the first class, could be MD simulations in the $NVT$ ensemble for liquid butane or bulk silicon, using well-defined standard force fields (e.g., CHARMM or Stillinger-Weber). Quantities of interest are typically computed during MD simulations (e.g., liquid densities). For flexibility, full trajectory files should also be stored
but some important observables might be worth precomputing (e.g., radial distribution functions).
The second class could include multi-billion atom MD simulations of dislocation formation \cite{Zepeda-Ruiz2017-sm} or solidification \cite{Miyoshi2019-ig,Shibuta2017-xl} or very long time-scale simulations of protein folding \cite{Piana2014-xy}.
For more complex use cases, the current infrastructure as discussed in Section ``Towards FAIR metadata schemas for computational materials science''
is not yet sufficient.
The challenges to be addressed are the need for support for (i) complex, heterogeneous, possibly multi-resolution systems; (ii) custom force fields; (iii) advanced sampling; 
(iv) classes of sampling besides MD (e.g., Monte Carlo, global structure prediction/search); (iv) larger simulations (i.e., need for sparsification of the stored data with minimal loss of information)

{\em Complex systems} include heterogeneous systems, e.g., adsorbate and surfaces, interfaces, solute (macro)molecules in solvent fluids, and multi-resolution systems, i.e., systems that are described at different granularity. The representation of complex systems requires a hierarchy of structural components, from atoms, through moieties, molecules, and larger (super)structures. Annotating such complexity will require human intervention as well as algorithm for automatically recognizing the structural elements (see, e.g., Ref. \cite{ARISE}).

Annotation of {\em force fields} into publicly accessible databases has been pioneered by OpenKIM \cite{tadmor2011potential} in materials science and MoSDeF \cite{mosdef} for soft matter. However, many simulations are performed with customized force fields. The field is already being augmented and will likely be further supported by machine-learning (ML) force fields. So far, the great majority of ML force fields are used only in the publication where they are defined. The {\em reusability}-oriented  annotation of force fields, including ML ones, require also establishing a criterion for comparing them. Comparisons can be carried out by means of standardized benchmark datasets, with a well-defined set of properties. Differences among predicted properties can establish a metric for the similarity of the force fields.

{\em Advanced sampling} techniques (e.g, metadynamics \cite{Laio2002-ss}, umbrella sampling\cite{Torrie1977-gi}, replica exchange\cite{Parisi}, transition-path \cite{dellago2002transition} and forward-flux \cite{Allen2009-mb} samplings) are typically supported by \textcolor{black}{libraries such as PLUMED \cite{Tribello2014-un} and OpenPathSampling \cite{Greff_da_Silveira2018-gy}. These libraries are used as plugins to codes where classical force-field-based (e.g. GROMACS \cite{abraham2015gromacs}, DL\_POLY \cite{todorov2006dl_poly_3}, LAMMPS \cite{thompson2022lammps}) or {\em ab initio} (e.g., CP2K \cite{kuhne2020cp2k} and Quantum Espresso \cite{giannozzi2009quantum}) MD, or both (e.g., i-Pi \cite{kapil2019pi}), are performed}. The input and output of these plugins will serve as the basis for the metadata related to these sampling techniques. In this regard, it would also be interesting to connect materials-science databases, such as the \mbox{NOMAD} Repository and Archive \cite{draxl2018nomad} or Materials Cloud Archive \cite{MaterialsCloud} to the PLUMED-NEST \cite{PLUMEDNEST}, the public repository of the PLUMED consortium\cite{theplumedconsortium_promoting_2019}, for example by allowing for automatic uploading of PLUMED input files to the PLUMED-NEST when uploading to the data repositories.

For \emph{long time-} and \emph{large length-scale} simulations, several questions arise: How should we deal with these simulations, where the extensive amount of data produced by MD simulations becomes overwhelmingly large to systematically store and share? Can we afford to store and share all of it? If the storage is limited or data retrieval is unpractically slow, how can we identify the significant and crucial part of the simulation to store it in a reduced form? Keeping the whole data locally and sharing the metadata with only the important parts of the simulations would be a viable alternative, assuming the different servers have enough redundancy. Standard analysis techniques such as similarity analysis and monitoring dynamics can also be used to identify the changes in structure and dynamics to store only the significant frames or specific regions in MD simulations (e.g., some QM/MM models uses large MM buffer-atom regions that may not be stored entirely). Further, the cost/benefit of storing versus running a new simulation must be weighed. On the other hand, researchers may soon face increased requirements from funding agencies to store their data for a number of years, in which case the present endeavour offers a convenient implementation. We note ongoing algorithmic developments on compression algorithms for trajectories, see, e.g., Ref. \cite{Brehm2018-un}.

\section*{Metadata for Computational Workflows}
\label{sec:workflow}

A computational workflow represents the coordinated execution of {\em repeatable} (computational) steps while accounting for {\em dependencies} and {\em concurrency} of tasks. In other words, a workflow can be thought as a script, a wrapper code that manages the scheduling of other codes, by controlling what should run in parallel, what sequentially and/or iteratively. This definition can be extended to workflows in experimental materials science or hybrid computational-experimental investigations, but, consistently with the previous sections, we limit the discussion to computational aspects only.

Once shared, workflows become useful building blocks that can be combined or modified for developing new ones. Furthermore, FAIR data can be reused as part of workflows completely unrelated to the workflows with which they were generated. An obvious example is artificial-intelligence-based data analytics, which can entail complex workflows involving data originally created for different purposes. 
During the last decade, the interest in workflow development has grown considerably in the scientific community\cite{Deelman2009} and various multi-purpose engines for managing calculation workflows, have been developed, including AFLOW\cite{aflow,aflowlib,aflowSTD}, AiiDA\cite{AiiDA-Pizzi2016,HMM2-Pizzi2018}, ASE\cite{larsen2017atomic}, and Fireworks\cite{Jain2015}. 
Using these infrastructures, a number of workflows have been used for scientific purposes, like convergence studies\cite{fleurconvergence-Broeder2018}, equations of state (e.g., AFLOW Automatic Gibbs Library \cite{Toher_PRB_AGL_2014} and the AiiDA common workflows ACWF\cite{huber2021common})
, phonons \cite{Phonopy,Petretto2018, Petretto2018a, 2D-Mounet2018}, elastic properties (e.g., the elastic-properties library for Inorganic Crystalline Compounds of the Materials Project\cite{MP_elastic}, AFLOW Automatic Elasticity Library, AEL\cite{AEL}, ElaStic\cite{golesorkhtabar2013elastic}), anharmonic properties (e.g., the Anharmonic Phonon Library, APL\cite{APL}, AFLOW Automatic Anharmonic Phonon Library, AAPL\cite{AAPL_2016}), high-throughput in the compositional space (e.g., AFLOW Partial Occupation, POCC \cite{aflowPOCC}), charge transport (e.g., organic semiconductors\cite{Symalla:2016dp,Friederich:2017ka}), of covalent organic frameworks (COFs) for gas storage applications \cite{mercado2018silico}, of spin-dynamics simulations\cite{russmann2021aiida}, high-throughput automated extraction of tight-binding Hamiltionians via Wannier functions\cite{valerio2020automated}, and high-throughput on-surface chemistry\cite{mishra2021observation}

There are two types of metadata associated to workflows. Thinking of a workflow as a code to be run, the first type of metadata characterizes the code itself. 
The second type is the annotation of a run of a workflow, i.e., its inputs and outputs. This type of metadata has been already described in Section ``Towards FAIR metadata schemas for computational materials science'',
together with a schematic list of possible workflow classes. It is important to realize that the inputs and outputs of the elementary-mode runs of the atomistic codes that are invoked in a workflow run are complemented by the inputs and outputs of the overarching workflows. A simple example: In an equation-of-state type of workflow, the energy and volume per unit cell of each single configuration that is part of the workflow is the output of the elementary run of the code, while the energy-vs-volume equation of state, e.g., fit to the Birch-Murnagham model, is an output of the workflow.

\section*{File Formats}
\label{sec:ff}
On an abstract level, a metadata schema is independent from its representation in computer memory, on a hard drive, or on just a piece of paper. But on a practical level, all data and metadata need to be managed, i.e., stored, indexed, accessed, shared, deleted, archived, etc.
File formats used in the community address different requirement and intended use cases. 
Some file formats privilege human readability (e.g., XML, JSON, YAML) but are not very storage efficient, others are binary and overall optimized for efficient searches, but require interpreters to be understood by a person (e.g., HDF5 \cite{folk2011overview}). 
There are a few use-cases and data properties in the domain of computational material science that are worth mentioning. First, such data are very heterogeneous and contain many simple properties (e.g., the name of a used code, or a list of considered atoms) that are mixed with properties in the form of large vectors, matrices, or tensors (e.g., the density of states or wave functions). The number of different properties requires hierarchical organization (e.g., with XML, JSON, YAML, or HDF5). It is desirable that many properties are easily human readable (e.g., to quickly verify the sanity of a piece of data), on the other hand large matrices should be stored as efficiently as possible for storing, retrieving and searching. Second, there are use cases where random (non-sequential) access of individual properties is desirable (e.g., return all band structures from a set of DFT calculations). 
Third, computational-material-science (meta)data need to be archived (efficient storage, prevention of corruption, backups, etc.) on one side, but they also need to be shared via APIs, e.g., for search queries. This requires to transform (meta)data from one representation in one file format (e.g., BagIt and HDF5) to another representation in a different format (e.g., JSON or XML).

These use-cases and data properties lead to the following conclusions. Even on a technical level, (meta)data need to be handled independent of the file format. Pieces of it have to be managed in different formats, and we need to be able to transform from one representation into another. Furthermore, if many different resources (files, databases, etc.) are used to store (meta)data from a logically conjoined dataset, references to these resources become an important piece of metadata itself. We propose to use an abstract interface (e.g., implemented as a Python library) based on an abstract schema.
This interface allows to manage (meta)data independent of the actual representation used underneath. Various implementations of such an abstract interface can then realize storage in various file formats and access to databases.

\section*{Metadata schemas for experimental materials science}
\label{sec:exp}
In contrast to computational materials science, in experimental materials science the atomic structure and composition is only approximately known. Several techniques are used to collect data that may be more or less directly interpreted in terms of the atomic and/or electronic structure of the material. In cases where the structure of the material is already known, careful characterization of properties helps to establish valuable relationships between structure and properties which, in turn, may help to refine theoretical models of these \textcolor{black}{structure-properties} links. The inherent uncertainty in every measurement process causes the precision with which data can be reproduced to be, in most cases, lower than in theoretical/computational materials science. These uncertainties are present even in a well-characterized experimental setup, i.e., when a comprehensive set of metadata is used.
In many cases it is not even the focus of an experiment to produce the most perfectly characterized data, but to invest just enough effort to address the specific question that drives the experiment.  

The information available about the material whose properties are to be measured is also much less complete than in the computational world, where often the position of every atom is known. However, while physical measurements may be limited in their precision, the accuracy with which a physically observable quantity is obtained is by definition of being physically observable much higher than in computational materials science, where the accuracy of the obtained physical quantity may depend strongly on the validity of approximations being made. 

The uncertainty in retrieving structure-property relationships in computational materials science, which depends on the suitability of the applied theoretical model and its computational implementation, translates in the real of experiments to an uncertainty in the atomic structure of the object that is being characterized and generally also some uncertainty in the measurement process itself. The metadata necessary to reproduce a given experimental data set must thus include detailed information about the material and its history together with all the parameters which are required to describe the state of the instrument used for the characterization. In most cases, both classes of metadata, i.e., those describing the material and those describing the instrument are going to be incomplete. While, for example, the full history of temperature, air pressure, humidity, and other relevant environmental parameters are not commonly tracked for the complete lifetime of a material (counter-examples exist, e.g., in pharmaceutical research), also information about the state of the instrument is not generally as comprehensive as it should ideally be (e.g., parameters are not recorded, or are not properly controlled, such as hysteresis effects in devices involving magnetic fields, or many mechanical setups). 

To overcome part of the uncertainty in the data, one needs to collect as many metadata about the material and its history, as possible, including those that one has no immediate use for at the moment, but might potentially need in the future. Since most of the research equipment being used for characterization tasks is commercial instrumentation, collecting this metadata in an (ideally) fully automated fashion requires the manufacturer’s support. In many cases the formats in which scientific data are provided by these instruments is proprietary. Even if all the data to describe the instrument’s condition of operation are stored, large parts of them may get lost when using the vendor's software to export the data to other formats; mostly because the ``standard format'' does not foresee storing vendor- and instrument-specific metadata. 
\textcolor{black}{It is however worth mentioning here that the CIF dictionaries (see section ``Towards FAIR metadata schemas for computational materials science''
) already contain (meta)data names to describe instrumentation, sample history, and standard uncertainties in both measured and computed values. As a useful addition, the CIF framework provides tools implementing quality criteria, which can be used for evaluating the trustworthiness of data objects. In this respect, the community has been developing with CIF a powerful tool onto which a FAIR representation of at least structural data can be built.} 

At large research infrastructures like synchrotrons and neutron scattering facilities, where a significant fraction of instruments is custom-built, and data are often shared with external partners, standards for file formats and metadata structures are being agreed upon the NeXus standard. NeXus \cite{nexusformat} defines hierarchies and rules on how metadata should be described and allows compliant storage using HDF5.
Experimental research communities can learn from these activities and provide {\em NeXus-format application definitions} which describe necessary metadata that should be stored in a dataset, along with definitions for some optional metadata. This common file format for scientific data is slowly beginning to spread to other communities. Having a standard file format for different types of scientific data seems to be an important step forward towards FAIR data management, since it severely reduces the threshold to share data across communities. Note that NeXus provides a glossary and connected ontology which helps in machine interpretability and so in {\em reusability}.

While standard file formats are of very high value in making data {\em findable} and {\em accessible}, due to common use of keywords to describe a given parameter, they also make them more {\em interoperable}, since the barrier for reading the data is lowered. However, making experimental data truly reproducible requires in many cases more metadata to be collected. Only if the uncertainty with which data can be reproduced is well understood, they may also be {\em reusable}. As discussed in the previous paragraph, part of these metadata must be provided by manufacturers of commercially available components of the experimental setup. 
Often this just requires more exhaustive data export functions and/or proper, i.e. versioned descriptions, for all of the instrument-state-describing metadata which are being collected during the experiment. Additionally, it may be necessary to equip home-built lab equipment with additional sensors and functionalities for logging their signals.

Even with added sensors and automated logging of all accessible metadata, in many cases, it is also necessary to compile and complete the record of metadata describing the current and past states of the sample that is being characterized by manually adding information and/or combining data from different sources.
Tools for doing this in a machine-readable fashion are Electronic Lab Notebooks (ELNs) and/or Laboratory Information Management Systems (LIMS). 
Many such systems are already available \cite{delageniere2011ispyb,malbet2013esrf,fisher2015synchweb,de2015ispyb,carp2017elabftw,bricogne2018achieving,tremouilhac2017chemotion}, including open-source solutions that combine features of both ELN and LIMS into one software. Server-client-solutions that do not require a specific client, but may be accessed through any web browser have the advantage that information may be accessed and edited from any electronic device capable of interacting with the server. Such ease of access, combined with the establishment of rules and practices of holistic metadata recording about sample conditions and experimental workflows will also help to increase the reproducibility and with that the {\em reusability} of experimental data. The easier the use of such a system is, and the more apparent it makes the benefits of the availability of FAIR experimental data, the faster it will be adopted by the scientific community.

\section*{Outlook on ontologies in materials science}
\label{sec:onto}

In data science, an ontology is a {\em formal representation of the knowledge of a community about a domain of interest, for a purpose}.
As ontologies are currently less common in basic materials science than in other fields of science, let us explain these terms:
\begin{itemize}
 \item {\em Formal representation} means that: 1) the ontology is a {\em representation}, \textcolor{black}{hence it is a simplification, or a model, of the target domain}, and 2) the attribute {\em formal} communicates that the ontological terms and relationships between them must have a deterministic and unambiguous meaning. Furthermore, {\em formal representation} implies that the mechanism to specify the ontology must have a degree of logical processing capability, e.g., inference and reasoning should be possible. Crucially, the attribute {\em formal} refers to the fact that an ontology should be machine readable.
 \item {\em Knowledge} is the accumulated facts, information, and skills of the experts of the domain of interest that are represented in the ontology. 
 \item The {\em community} influences the ontology in two aspects; 1) it implies an overall agreement between a group of experts/users of the knowledge as represented in the ontology and 2) it indicates that the ontology is not meant to convince a whole population nor wants to be universal. However, if it fulfills the requirements of bigger communities, the ontology will be adopted by broader audiences and will find its way towards standardization. 
 \item The {\em domain of interest} is the common ground for the community, e.g., a scientific discipline, a subordinate of discipline, or a market section. It is often used as a boundary to limit the scope of the ontology. It is a proper tool to detect overlapping concepts, modularizing ontologies, and identifying extension and integration points. 
 \item The {\em purpose} conveys the goals of the ontology designers so that the ontology is applicable to a set of situations. In many ontology design efforts, the purpose is formulated by a collection of so-called {\em competency questions}. These questions and the answers provided to them identify the intent and viewpoint of the designers and set the potential applications of the ontology.\\
\end{itemize}

In practice, ontologies are often mapped onto, and visualized by means of, directed acyclic graphs, where an edge is one of a well-defined set of relationships (e.g., {\em is a}, {\em has property}) and each node is a {\em class}, i.e., a concept which is specific to the domain of interest. Each node-edge-node {\em triple} is interpreted as a subject-predicate-object expression. For instance, in an ontology for catalysis, one could find the triples: ``catalytic material -- has property -- selectivity'', and ``selectivity -- refers to -- reaction product''. 
Ontologies address the {\em interoperability} requirement of FAIR data. By means of a machine-readable formal structure, which can be connected to an existing or {\em ex novo} derived metadata schema of a database, ontologies allow queries over various databases, even from different fields.

The literature already contains several ontologies created for representing (aspects of) materials science. The most ambitious project is probably EMMO\cite{cemmo2020}, which stands for both European Materials Modelling Ontology, developed within the European Materials Modelling Council (EMMC), and Elemental Multiperspective Material Ontology. EMMO is designed to provide a formal way to describe the fundamental concepts of physics, chemistry, and materials science, to provide an all-purposes common ground for describing materials, models, and data that can be adapted by all sub-domains of condensed-matter physics and chemistry. The development of EMMO includes also a handful of {\em domain ontologies} that assume EMMO as top-level ontology\cite{emmo-repo}. These domain ontologies span subjects such as ``atomistic and electronic modeling'', ``crystallography'', ``mechanical testing'', and more. So far, however, EMMO and its domain ontologies have not been connected to existing databases.

Other domain-specific ontologies, not related to EMMO, have been developed. For instance, the Materials Ontology\cite{ashino2010materials} was developed for the exchange of data among databases for thermal properties, the MatOnto ontology \cite{cheung2008towards} addresses oxygen ion conducting materials in the fuel cell domain, the NanoParticle Ontology \cite{thomas2011nanoparticle} maps properties of nanoparticles with the purpose of designing new nanoparticles with given properties, while the eNanoMapper ontology \cite{hastings2015enanomapper} focuses on assessing risks related to the use of nanomaterials from the engineering point of view. 

An application-oriented ontology is Materials Design Ontology (MDO) \cite{li2020ontology}, developed under the guidance of the schemas from OPTIMADE\cite{Andersen2021}, and therefore aimed at dealing with data from the various materials data repositories (AFLOW, Materials Project, etc.) on a common ground. In practice, MDO connects calculated structures with the calculated properties and the  physical model adopted to calculate structures and properties. Furthermore, the provenance for each calculation, is also represented in MDO. It has recently been extended using text mining on thousands of journal articles \cite{li2019method}.

The hierarchical structure of {\em \mbox{NOMAD} Metainfo} already includes ontological aspect. More specifically, it represents atomistic calculations, as performed by all the parsed simulation codes. {\em \mbox{NOMAD} Metainfo} contains already five types of relations between the metadata: (a) is subclass of, (b) is part of, (c) has reference, (d) has dimension and (e) has category. The latter relation, {\em has category} is introduced to describe conceptually physical quantities (e.g., ``energy'', ``velocity'', etc.).
Recently \cite{Maja_thesis}, this basic {\em \mbox{NOMAD} Metainfo} ontology has been expanded to include a representation of operations among arrays (in an ontology, any mathematical concept needs to be represented in order to properly operate with the physical quantities in complex queries). This extension allowed for the introduction of the notion of ``similarity'' relationship that has been applied as a proof of concept to the calculated electronic density of states, as stored in the \mbox{NOMAD} Archive, in order to identify materials with similar electronic structures \cite{kuban2022density,kuban2022similarity}.

\section*{Conclusions and Outlook}
Defining --- as completely as possible --- a pool of metadata for all the methods and computed quantities described above, is crucial for processing, storing, and providing FAIR materials-science data. A key challenge is the mapping into a metadata schema of the full set of input parameters, including those hidden into the specific codes, and all the available output. This practice will facilitate reproducibility, benchmarking, and peer-review processes. 

In particular, we emphasize the importance of developing a hierarchical and modular metadata schema in order to represent the complexity of materials science data and allow for access, reproduction, and repurposing of data, from single-structure calculations to complex workflows. Furthermore, the modularity of the schema enables its extensibility, which is vital for the long-term maintenance of the metadata infrastructure. 

As an example, we presented the current status of the \mbox{NOMAD} metadata schema, which was designed to comply with the FAIR principles. By means of existing parsers that map a growing set of atomistic-simulation code packages into the hierarchical, modular \mbox{NOMAD} metadata schema, the \mbox{NOMAD} infrastructure already provides the community with a FAIR storage of materials science data. The challenges of fully covering the ground-state electronic calculations, and extending the schema to excited states, dynamical simulations, and complex workflows were examined in detail. By means of a community effort, all aspects of the different subfields, and all the practical details of each specific implementation can be mapped on the \mbox{NOMAD} metadata schema. 
Finally, we discussed the challenges of the {\em FAIRification} of experimental materials-science metadata and the creation of ontologies for materials science. Ontologies will unlock the {\em interoperability} of the FAIR data by enabling the access and reuse of data across materials-science areas, but also outside materials science.\\
\textcolor{black}{As a perspective, probably the biggest benefit of meeting the interoperability challenge will be to allow for routine comparisons between computational evaluations and experimental observations. In fact, it is not trivial to associate a given computed quantity, derived through a given theoretical modelling, to an experimentally measured quantity. This association requires the judgment of a domain expert and a full characterization of both compared quantities. This is where a formalized ontology, applied to FAIR data in materials science, could automatize the process.}




\section*{Acknowledgements} 
We would like to thank all the participants to the workshop ``Shared Metadata and Data Formats for Big-Data Driven Materials Science: A \mbox{NOMAD}--FAIR-DI Workshop'', as listed at \url{https://th.fhi-berlin.mpg.de/meetings/META2019/}, who have contributed with questions and comments to ideas discussed in this paper.

The organizers of and participants to the OMDI2021 workshop (see \url{https://liu.se/en/research/omdi2021} for the full list of names) are acknowledged for insightful discussions that inspired some of the concepts discussed in Section ``Outlook on ontologies in materials science''.

This work received funding by the European Union’s Horizon 2020 research and innovation program under the grant agreement Nº 951786 (\mbox{NOMAD} CoE) and by the German Research Foundation (DFG) through the NFDI consortium FAIRmat, project 460197019.

\section*{Author contributions}
The present paper is inspired by and based on the minutes of the work-groups discussions at the workshop ``Shared Metadata and Data Formats for Big-Data Driven Materials Science: A \mbox{NOMAD}--FAIR-DI Workshop''. Here, we report the composition of the original work groups, which reflect into the main contributions to the paper's sections.
{\bf Metadata, metadata schemas and ontologies} (Introduction, Section ``Towards FAIR metadata schemas for computational materials science'' and section ``Outlook on ontologies in materials science'':
Patrick Lambrix, Javad Chamanara, Carsten Baldauf, Tatyana Sheveleva, Benjamin Regler, Alvin Noe Ladines, Christoph T. Koch, Christof W\"{o}ll, Stefano Cozzini, Astrid Schneidewind, Maja-Olivia Lenz-Himmer; 
{\bf Ground-state calculations} (Section ``Metadata for ground-state electronic-structure calculations''
): Micael Oliveira, Sergey Levchenko; 
{\bf Perturbative and excited-states calculations} (Section ``Metadata for external-perturbation and excited-state electronic-structure calculations''
): Claudia Draxl, Pasquale Pavone, Denis Usvyat; 
{\bf Potential-energy sampling} (Section ``Metadata for potential-energy sampling''
): James Kermode, Tristan Bereau, Christian Carbogno, Omar Valsson, Markus Kühbach, Chuanxun Su, Ron Miller, Berk Onat; 
{\bf Workflows} (Section ``Metadata for Computational Workflows''
): Stefano Curtarolo, Shyam Dwaraknath, Adam Michalchuk, Giovanni Pizzi, Gian-Marco Rignanese, Jörg Schaarschmidt; 
{\bf Data formats} (Section ``File Formats''
): Ádám Fekete, Markus Scheidgen; 
{\bf Metadata for experiments} (Section ``Metadata schemas for experimental materials science''
): Christoph T. Koch, Sandor Brockhauser, Astrid Schneidewind.\\
Luca M. Ghiringhelli and Matthias Scheffler coordinated the formation of the work groups, participated to the discussions in several work groups, and prepared the first draft of the paper.  All authors contributed to the final version of the paper.


%

\flushbottom

\end{document}